\documentclass[lnbip]{svmultln}
\usepackage{graphicx}

\newcommand{\bpdmnapp}[1]{\fbox{\textbf{#1}}}
\newcommand{\bpdmndata}[1]{\underline{\textbf{#1}}}

\begin{document}
\frontmatter          
\pagestyle{headings}  

\mainmatter
%
%
\title{\texttt{ChOrDa}: a methodology for the modeling of business processes with BPMN}
\titlerunning{}  
%
\author{Matteo Buferli\inst{1} \and Matteo Magnani\inst{1} \and Danilo Montesi\inst{1}}
\authorrunning{Buferli et al.}   
%
\tocauthor{Matteo Buferli (University of Bologna),
Matteo Magnani (University of Bologna),
Danilo Montesi (University of Bologna)}
\institute{Department of Computer Science,\\
University of Bologna\\
\email{{buferli,magnanim,montesi}@cs.unibo.it}}

\maketitle              


\begin{abstract}
In this paper we present a modeling methodology for BPMN, the standard notation for the representation of business processes.
Our methodology simplifies the development of collaborative BPMN diagrams, enabling the automated creation of skeleton process diagrams representing complex choreographies.
To evaluate and tune the methodology, we have developed a tool supporting it, that we apply to the modeling of an international patenting process as a working example.
\end{abstract}


\section{Introduction}
\label{intro}

While it is typical to deal with business processes that start and end inside a single organization, many processes are not constrained inside the walls of a single company. For example, supply chains can be seen as large processes involving suppliers, manufacturers, distributors and retailers. Unfortunately, processes involving many actors may easily become very complex, and this complexity may be inherited by the tools (languages/notations and editors) used to design or document them.

BPMN is the OMG standard notation for the representation of business processes \cite{BPMN}. Even if BPMN constructs are very intuitive, large business process diagrams involving both collaborations between several actors and details on the single participants can be very hard to design or document without any associated methodology --- much like building a miniature ship model with glue and screws, but without instructions. In this paper we introduce a methodology for the translation of informal process descriptions into BPMN diagrams, and a software design tool supporting it. In particular, we provide a tool to automatically generate diagrams showing the collaborative aspects of the process, starting from annotated textual requirements.

Why do we need a specific methodology for BPMN? This lies in the very nature of this language, which enables the representation of three important aspects of business processes, making it a unique modeling tool. First, we can represent a \emph{choreography} of processes, i.e., how different processes interact with each other to fulfill a common objective. Second, we can represent the \emph{orchestration} of a process, i.e., its internal organization into sequences of activities. Third, BPMN allows the representation of the same information at different levels of detail, using sub-processes --- this being fundamental to provide different views of the same process to people with different roles, like top managers and technical staff.

\begin{figure}
\begin{center}
\includegraphics[width=\textwidth]{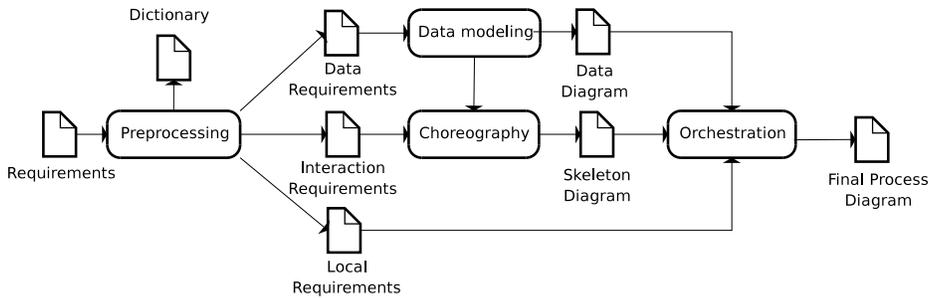}
\end{center}
\caption{A diagram summarizing the proposed methodology}
\label{fig:methodology}
\end{figure}

In summary, BPMN enables the representation of complex scenarios because it can include many different aspects into a single diagram: \emph{\underline{ch}oreography}, \emph{\underline{or}chestration}, and \emph{\underline{da}ta}, at several different \emph{levels of abstraction} \cite{BPMN,Barros06}. Therefore, the main idea behind our approach is that the initial requirements can be split into different classes, that can be specifically addressed during well separated and thus simplified modeling steps. As we have illustrated in Figure~\ref{fig:methodology}, after a typical pre-processing of the available informal requirements aimed at removing ambiguities and producing a dictionary with all definitions and synonyms, we split them into atomic statements referring to one of the following aspects: (D) data, (I) interactions between different participants, and (L) local work of a single participant. At this point, each class can be processed independently from the others.

To the best of our knowledge, this is the first presentation of a methodology for BPMN modeling, and in particular of a methodology to automate the drawing of the collaborative portion of distributed processes. Obviously, it is based on best practices taken from existing data and software modeling methodologies like the IBM Rational Unified Process (\texttt{www-01.ibm.com/software/awdtools/rup}), the IDEF methods (\texttt{http://www.idef.com}), data modeling using ER diagrams and Object Process Modeling (\texttt{http://www.objectprocess.org}). The POEM (Process Oriented Enterprise Modelling) methodology uses BPMN as one of several basic diagram types. Although we are not aware of existing presentations of this methodology, still under development, it seems to have a wider scope than our proposal, covering several additional aspects of an enterprise, while we present specific results regarding the BPMN notation.

The design tool developed to support our methodology is a plug-in for Microsoft Visio\footnote{http://office.microsoft.com/visio}, and uses an extension of the POEM stencils\footnote{http://bpmnpop.sourceforge.net}. The tool can be used to design BPMN diagrams, to annotate them with additional attributes (like the cost of activities) and to generate their XPDL representations \cite{MagnaniBPM07,XPDL}. A time limited beta version of the tool can be downloaded from the Web site http://bpm.cs.unibo.it.

\subsection{An Overview of the Methodology: Main Phases}

\begin{figure}
\includegraphics[width=\textwidth]{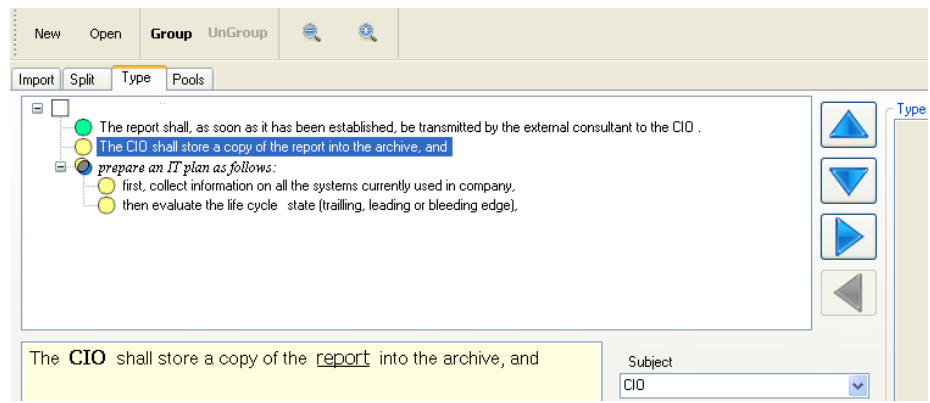}
\caption{Requirements managed using our design tool}
\label{fig:requirements}
\end{figure}

First, we need to split the requirements into small atomic statements, and assign each statement to one of the three aforementioned classes. The user interface to import, edit and annotate the requirements is illustrated in Figure~\ref{fig:requirements}. After the identification of data and participants, that can be tagged directly on the imported text using our tool, the assignment will usually be straightforward:
\begin{itemize}
\item If the statement concerns only data, then it will be a \textbf{Data} requirement.
\item If the statement contains one single participant (and describes one or more actions), then it will be a \textbf{Local} requirement.
\item If the statement refers to two participants, this will indicate an \textbf{Interaction} requirement, and we should also be able to identify the exchanged data.
\end{itemize}

\subsubsection{Data modeling} We can now start modeling the data (D), which is a primary component of real business processes and will thus drive all the modeling activities. In fact, a process is basically a sequence of activities aimed at modifying some data or objects, and the production of new data is the way in which business processes generate value --- for example, many business processes are used to transform raw materials into final products. As BPMN offers a limited support to the modeling of data, being more activity/event-oriented, these requirements can be addressed by existing data modeling approaches and attached to the diagrams as complementary documentation. In our tool we use an extended version of BPMN, obtained by including some features of ER and data flow diagrams \cite{Chen76,Gane77,Yourdon06}. However, this is not required by our methodology, which applies to standard BPMN diagrams, therefore we will not present the details of our extension in this work.

After this modeling step, we will have identified a list of all the data/objects referenced in the requirements --- the next step will be the definition of how they are exchanged among different participants.

\begin{figure}
\begin{center}
\includegraphics[width=.45\textwidth]{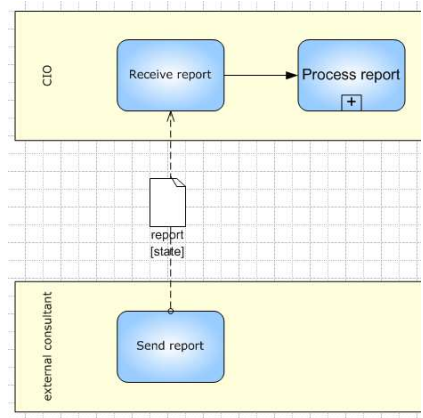}
\end{center}
\caption{The skeleton diagram corresponding to the statement: \emph{The \underline{report} shall, as soon as it has been established, be transmitted by the \emph{\textbf{external consultant}} to the \emph{\textbf{CIO}}}. This diagram has been generated automatically, starting from this annotated statement}
\label{fig:choreography}
\end{figure}

\subsubsection{Interaction modeling} Data flows will then be used to generate a so called \emph{skeleton diagram} representing how data is exchanged between the participants to produce the final products of the process. Basically, during this step we focus on choreography, i.e., we identify all the participants and their interactions (I). Each participant is represented using a BPMN pool, and we draw a message flow between two of them for each requirement. In this way, after having identified all interaction requirements we can automatically build a skeleton diagram, as we have exemplified in Figure~\ref{fig:choreography}: each interaction between participants A and B corresponds to a \texttt{Send Data} activity in A, a \texttt{Receive Data} activity in B, and a \texttt{Process Data} sub-process in B, indicating that the received data will be later manipulated --- this will be expanded during the next step of the methodology. Notice that in this paper we do not deal with the verification of choreographies, but with the formalization of existing informal descriptions of a choreography --- automated verification tools will obviously be of great utility to check the designed diagram, but this is an orthogonal problem with respect to the scope of this paper.

\subsubsection{Local modeling} 
\begin{figure}
\begin{center}
\includegraphics[width=\textwidth]{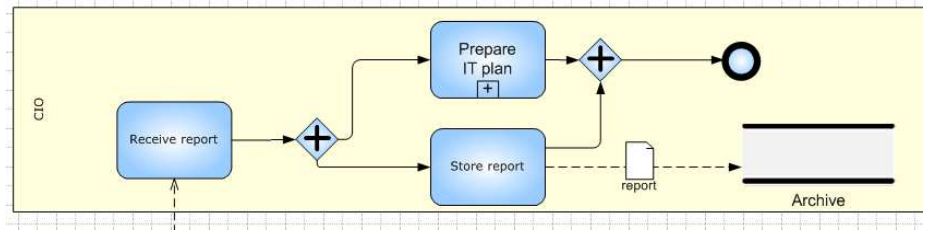}
\end{center}
\caption{The modeling of local (L) requirements, in a top-down fashion}
\label{fig:orchestration}
\end{figure}

Now, as we exemplify in Section~\ref{example}, we will have a skeleton diagram with all the participants (pools) and all messages exchanged between them, representing the complete (abstract) data paths used to produce the final outcome of the process, be it a document, a product, or any other valuable object. For each exchanged message, we will also have a sub-process (the rectangle with a small \emph{plus} sign represented in Figure~\ref{fig:choreography}) hiding the local activities performed by the participant to manipulate the data. Therefore, we can focus on the remaining requirements (L) describing these activities. This modeling step can be performed in a top-down way, following the philosophy behind BPMN which uses abstraction levels as a basic tool to provide different views on the same process. Therefore, L-statements will be associated to specific sub-processes, hierarchically organized, and finally added to the diagram. For example, consider the following statements:
\begin{enumerate}
\item The CIO shall store a copy of the report into the archive, and
\item prepare an IT plan as follows:
\begin{enumerate}
\item first, collect information on all the systems currently used in the company,
\item then evaluate their life cycle state (trailing, leading or bleeding edge).
\end{enumerate}
\end{enumerate}
These can be modeled as illustrated in Figure~\ref{fig:orchestration}, where we have also used a \emph{store} (the archive) that is one of our data modeling extensions, borrowed from Data Flow Diagrams \cite{Gane77}, and whose meaning should be intuitively evident. Later, the \texttt{Prepare IT plan} sub-process can be expanded including the statements describing this activity (2.a, 2.b and following, in this example), and so on recursively.

\section{Working Example}
\label{example}

In this section we apply our methodology to a summarized version of the process described in the Patent Cooperation Treaty\footnote{http://www.wipo.int/pct/en/texts/articles/atoc.htm}, providing instructions on how to file an international application for a patent that can be then submitted to the patent offices of some of the States covered by the treaty.

\subsection{Participants, Data and Glossary}

We start our example by assuming to work on a single text file. This file can be processed
as usual, by clarifying unclear sentences, identifying synonyms, replacing them with
consistent terms, and building a technical dictionary.

\begin{figure}
\fbox{\begin{minipage}{\textwidth}
 \bpdmndata{Applications} for the protection of inventions in any of the
Contracting States may be filed as \bpdmndata{international applications}.
 An \bpdmndata{international application} shall contain a request, a description,
one or more claims, one or more drawings (where required), and an abstract.
 The \bpdmndata{request} shall contain: the designation of the Contracting States in which
protection for the invention is desired; the name of and other prescribed data concerning
the \bpdmnapp{applicant}; the title of the invention; the name of and other prescribed data concerning the inventor.
The \bpdmnapp{receiving Office} shall accord as the international filing date
the date of receipt of the \bpdmndata{international application},
 provided that it has verified some basic requirements.
 If the \bpdmnapp{receiving Office} finds that the international application
did not, at the time of receipt, fulfill these requirements, it shall invite
the \bpdmnapp{applicant} to file the required correction.
If the \bpdmnapp{applicant} complies with the invitation,
 the \bpdmnapp{receiving Office} shall accord as the international filing date the date
of receipt of the required \bpdmndata{correction}.
 One copy of the \bpdmndata{international application} shall be kept by the
\bpdmnapp{receiving Office} (\bpdmndata{home copy}),
 one copy (\bpdmndata{record copy}) shall be
transmitted to the \bpdmnapp{International Bureau}, and
 another copy (\bpdmndata{search copy})
shall be transmitted to the competent \bpdmnapp{International Searching Authority}.
 International search shall be carried out by an \bpdmnapp{International
Searching Authority}, whose tasks include the establishing of documentary search reports on
prior art with respect to inventions which are the subject of applications.
 The \bpdmndata{international search report} shall, as soon as it has been
established, be transmitted by the \bpdmnapp{International Searching Authority} to
 the \bpdmnapp{applicant} and
 the \bpdmnapp{International Bureau}.
 The \bpdmnapp{applicant} shall, after having received the \bpdmndata{international search
report}, be entitled to one opportunity to amend the claims of the international
application by filing \bpdmndata{amendments} with the \bpdmnapp{International Bureau} within the
prescribed time limit.
 The \bpdmnapp{International Bureau} shall publish \bpdmndata{international applications}
promptly after the expiration of 18 months from the priority
date of that application.
 The \bpdmnapp{applicant} shall furnish a copy of the \bpdmndata{international search
report}, of the \bpdmndata{international application} and a \bpdmndata{translation} thereof, and pay the national fee, to each
\bpdmnapp{designated Office} not later than at the expiration of 30 months
from the priority date.
\end{minipage}}
\caption{Working example: international patent application filing procedure}
\label{fig:txt}
\end{figure}

Then, we may identify all the data objects mentioned in the requirements, and all the participants. In Figure~\ref{fig:txt} we have represented the requirements describing this example, already formatted according to this preliminary analysis. In particular, we have identified and underlined all data objects:
\begin{itemize}
\item application,
\item international application,
\item request,
\item home copy,
\item record copy,
\item search copy,
\item international search report, and
\item translation.
\end{itemize}
In addition, we have identified and boxed all participants:
\begin{itemize}
\item applicant,
\item receiving office,
\item International Bureau,
\item Internation Searching Authority, and
\item designated office.
\end{itemize}

At this point, we may easily proceed by splitting the requirements into three groups. This can be done using our tool, as we have illustrated in Figure~\ref{fig:requirements} on the examples illustrated in the first part of the paper, where we can import the requirements, split them into statements and assign each statement to the \emph{data} class, to the \emph{interaction} class (statements where two participants are involved) and to the \emph{local} class (statements describing the activity of a single participant). Notice that we can also group several L-statements together, assigning a name to the group, and do so recursively --- in this way we can support a top-down modeling approach for the orchestration step, discussed later.
The next three activities will consist in the manipulation of these three sets of requirements.

\subsection{Data}

First, we model the data that will be manipulated by the process. As we have aforementioned, we do not provide details of this phase, that is performed according to existing data modeling methodologies. For example, an international application (one of the data objects in the working example) is composed of several parts, like a request and a description, that can be visualized on the final diagrams using our tool or illustrated in other diagrams using specific notations.


\subsection{Choreography}

\begin{figure}
\mbox{}\hspace{-55pt}\includegraphics[height=22cm]{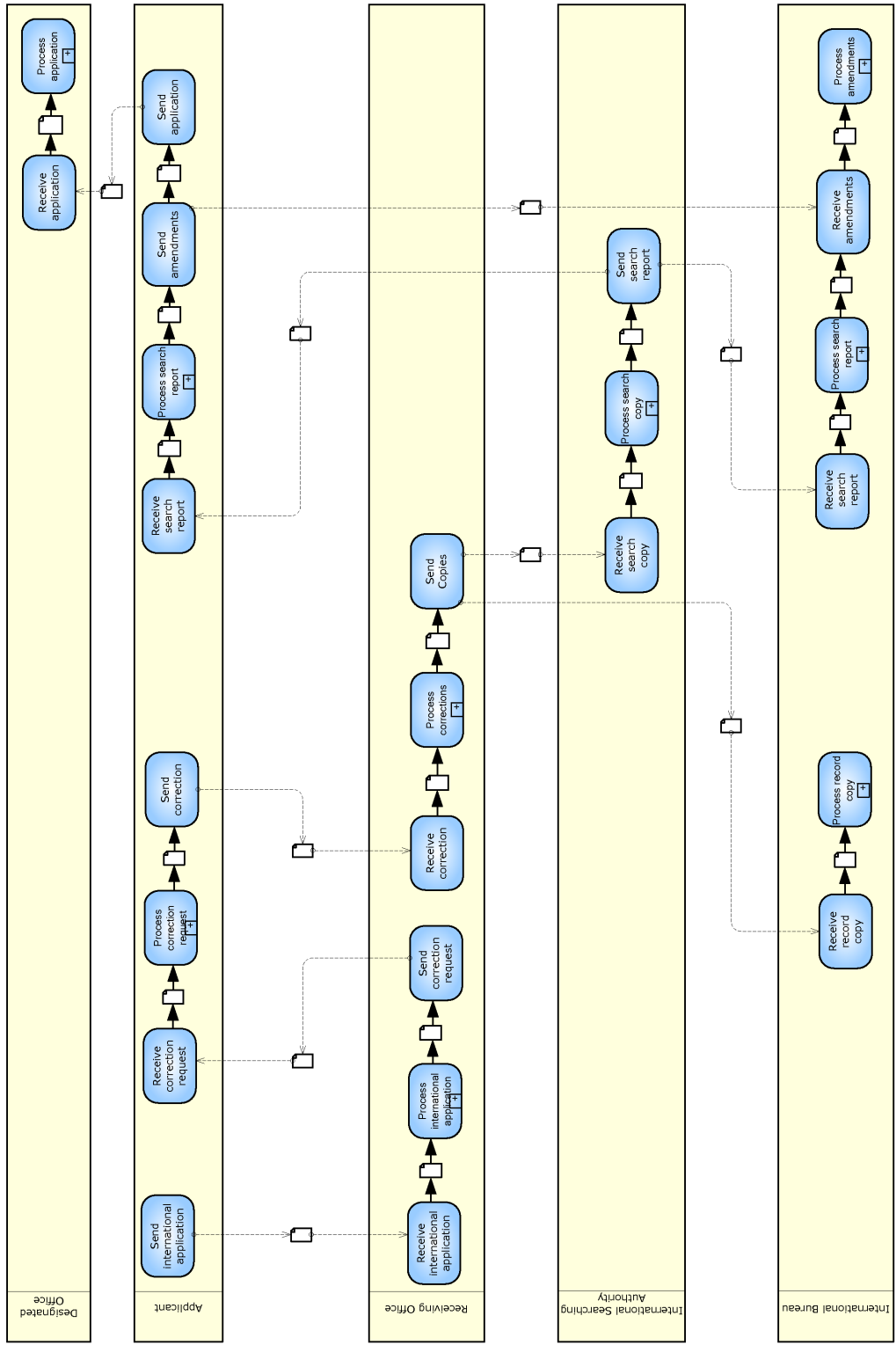}
\caption{Process Skeleton obtained using our tool. Also this diagram has been generated automatically, starting from the annotated textual description of the process}
\label{fig:skeleton}
\end{figure}

Now, we can start describing the dynamics of the data, focusing on data exchange between different participants. Using the approach described in the introduction, we can translate each I-statement into a triple of activities, generating the diagram represented in Figure~\ref{fig:skeleton}. This diagram, which derives directly from the isolated statements and can be thus easily produced semi-automatically, clearly describes the main data paths and gives a first, high level view of the process.

\subsection{Orchestration}

\begin{figure}
\mbox{}\hspace{-55pt}\includegraphics[height=22cm]{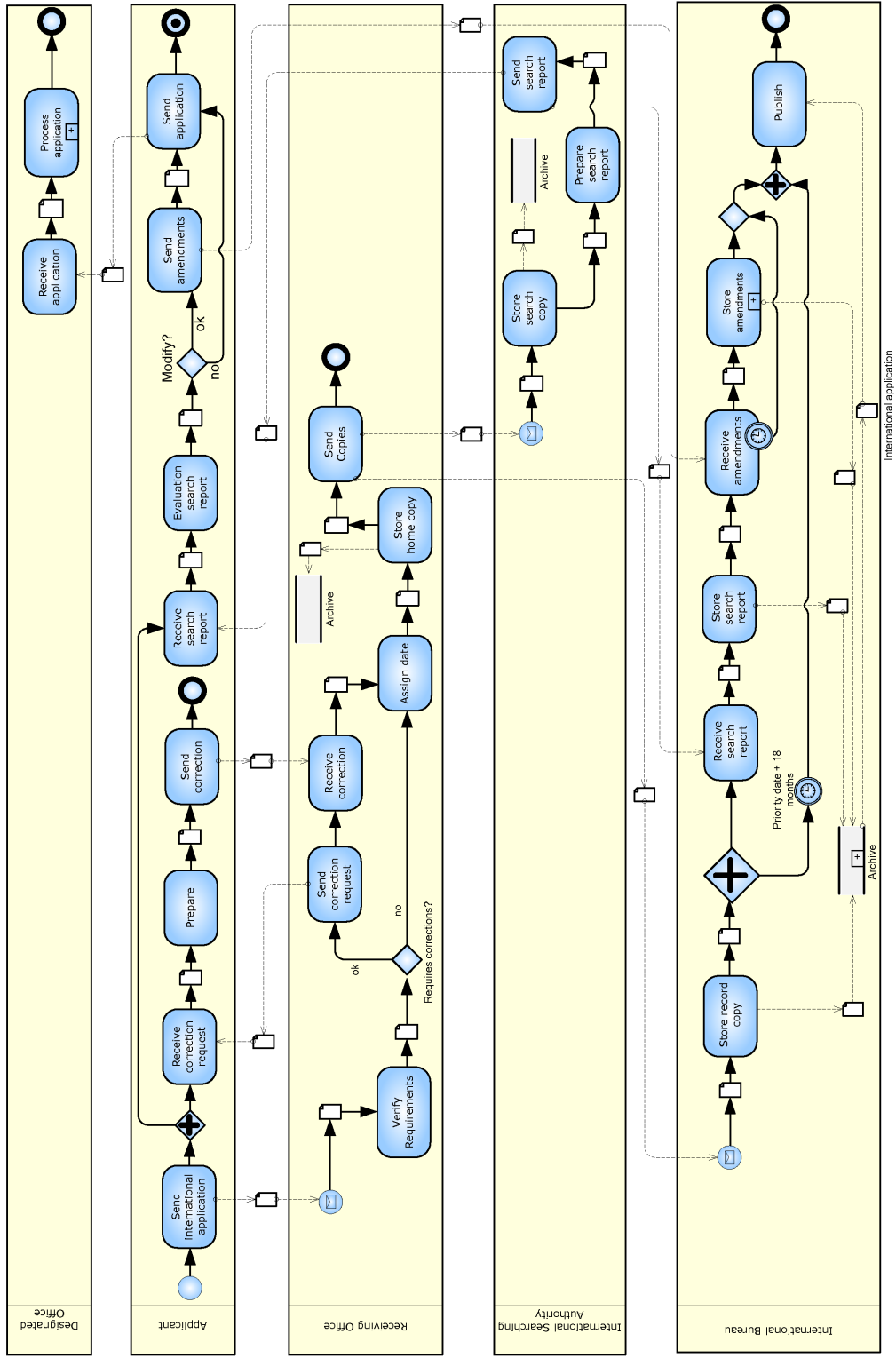}
\caption{Final process}
\label{fig:final}
\end{figure}

At this point, we can refine the skeleton diagram including the details of each participant (pool) in a top-down way, associating each local statement from the requirements to one specific sub-process. For instance, let us focus on the \texttt{Process International Application} sub-process in the \texttt{Receiving Office} pool\footnote{We can also consider different pools as lanes of a single pool, but this is not relevant to our discussion}.
There are three statements corresponding to this activity:
\begin{itemize}
\item The receiving Office shall accord as the international filing date
the date of receipt of the international application,
\item provided that it has verified some basic requirements.
\item the receiving Office shall accord as the international filing date the date
of receipt of the required correction.
\end{itemize}
The modeling of these requirements is extremely easy, and this derives from the fact that we can model separately each sub-process in the skeleton, and focus on a limited portion of the final diagram, thanks to our previous modeling steps and statement splitting. This can be repeated for each sub-process, leading to the definition of the complete diagram illustrated in Figure~\ref{fig:final}.

\section{A Summary of the Methodology}
\label{methodology}

The methodology proposed in this paper is composed of the following main steps:
\begin{enumerate}
\item Pre-process the initial requirements (remove ambiguities, update and refine unclear sentences and generate a dictionary with explanations of the technical terms and indications of synonyms).
\item Split the requirements into elementary statements.
\item Separate data (D) statements from activity statements.
\item Identify the participants, and mark each statement as a local (L) activity (orchestration) or an interaction (I) among participants (choreography).
\item Draw the skeleton of the process, modeling interaction activities.
\item Tree-structure local activities, associate them to sub-processes in the skeleton diagram, and model them in a top-down way by increasing the level of detail at each iteration (if necessary).
\end{enumerate}

In this way, part of the modeling activity can be semi-automated, and the definition of the orchestration inside each pool can be performed by focusing on small portions of the initial requirements. In addition, each statement can be easily associated to a specific part of the final diagram, and it can be then verified if the diagram is \emph{complete}, i.e., if it models all the requirements.

To the best of our knowledge, this work describes the first methodology to support the modeling of collaborative business processes with BPMN, and has been developed focusing on the specific features of this notation. In addition, using the tool exemplified in this paper we have been able to evaluate this methodology and to appreciate the simplification that it enables, and that has been described and highlighted on the working example presented in this paper.

%

%

%
%

\end{document}